\documentclass[twocolumn,english,prl,amssymb,aps,superscriptaddress,showpacs,twocolumn,amsmath,showkeys,floatfix]{revtex4-2}

\usepackage{geometry}
\usepackage[active]{srcltx}
\usepackage{textcomp}
\usepackage{amsmath}
\usepackage{graphicx}
\usepackage{color}
\usepackage{esint}

\makeatletter

\usepackage[english]{babel}
\usepackage{physics}
\usepackage[breaklinks]{hyperref}
\usepackage{inputenc}

\makeatother

\begin{document}

\title{Dissipative Kerr solitons, breathers and chimera states in coherently driven passive cavities with parabolic potential}

\author{Yifan Sun}\email{yifan.sun@uniroma1.it}
\affiliation{Department of Information Engineering, Electronics and Telecommunications, Sapienza University of Rome, Via Eudossiana 18, 00184 Rome, Italy}
\author{Pedro Parra-Rivas}
\affiliation{Department of Information Engineering, Electronics and Telecommunications, Sapienza University of Rome, Via Eudossiana 18, 00184 Rome, Italy}
\author{Mario Ferraro}
\affiliation{Department of Information Engineering, Electronics and Telecommunications, Sapienza University of Rome, Via Eudossiana 18, 00184 Rome, Italy}
\author{Fabio Mangini}
\affiliation{Department of Information Engineering, Electronics and Telecommunications, Sapienza University of Rome, Via Eudossiana 18, 00184 Rome, Italy}
\author{Mario Zitelli}
\affiliation{Department of Information Engineering, Electronics and Telecommunications, Sapienza University of Rome, Via Eudossiana 18, 00184 Rome, Italy}
\author{Raphael Jauberteau}
\affiliation{Department of Information Engineering, Electronics and Telecommunications, Sapienza University of Rome, Via Eudossiana 18, 00184 Rome, Italy}
\author{Francesco Rinaldo Talenti}
\affiliation{Department of Information Engineering, Electronics and Telecommunications, Sapienza University of Rome, Via Eudossiana 18, 00184 Rome, Italy}
\author{Stefan Wabnitz}
\affiliation{Department of Information Engineering, Electronics and Telecommunications, Sapienza University of Rome, Via Eudossiana 18, 00184 Rome, Italy}
\affiliation{CNR-INO, Istituto Nazionale di Ottica, Via Campi Flegrei 34, 80078 Pozzuoli, Italy}

\begin{abstract} 
    We analyze the stability and dynamics of Kerr dissipative solitons (DKSs) in the presence of a parabolic potential. This potential stabilizes oscillatory and chaotic regimes, favoring the generation of static DKSs. Furthermore, the potential induces the emergence of new dissipative structures, such as asymmetric breathers and chimera-like states. Based on a mode decomposition of the previous dynamics, we unveil the underlying modal interactions.  
\end{abstract}

\maketitle

Dissipative temporal Kerr soliton (DKS) \cite{wabnitz:93} generation and manipulation have been an emerging topic in photonics over the past decade, since they provide a breakthrough framework for coherent frequency comb generation in chip-scale microresonator platforms \cite{herr_temporal_2014-1,PASQUAZI20181}. In contrast to conservative systems, where solitons are formed due to a counter-balance between dispersion and nonlinearity, dissipative solitons additionally require an equilibrium between internal dissipation and external energy flow or driving. The dynamics and stability of DKSs have been analyzed in detail in the mean-field approximation, where passive Kerr resonators are described by a driven and damped nonlinear Schroedinger model \cite{haelterman_dissipative_1992,chembo_theory_2017}. In this context, a large variety of DKSs emerge in anomalous and normal dispersion regimes \cite{parra-rivas_dark_2016,parra-rivas_bifurcation_2018-1,parra-rivas_origin_2021}. As the pump intensity grows larger, DKSs undergo different types of instabilities, leading to complex spatio-temporal dynamics, which can be either periodic (i.e., breathers) or chaotic \cite{leo_temporal_2010,anderson_observations_2016,liu_characterization_2017,lucas_breathing_2017,bao_observation_2018}. 

Spatio-temporal dynamics can be stabilized through 
high-order effects, such as third-order dispersion, which considerably reduces the extension of unstable parameter regions in favor of static DKSs \cite{milian_soliton_2014,parra-rivas_third-order_2014}. Moreover, third- and fourth-order dispersion effects may lead to the appearance of new types of localized states, and to the coexistence of bright and dark DKSs \cite{tlidi_high-order_2010,parra-rivas_coexistence_2017,li_experimental_2020}, as it also does the Raman effect \cite{parra-rivas_influence_2020}.  
Spatio-temporal instabilities may also be suppressed by the modulation of the intracavity background field. These modulated defects can be induced 
through the external phase of the driving field \cite{Wabnitz:96,Jang2015,Cole2018,Hendry2019,Talenti2020,Erkintalo2022}, or 
by synchronous intracavity phase modulation \cite{Mecozzi:92,Englebert2021}. The latter can be introduced via an electrooptic modulator, and it leads to a synthetic dimension \cite{Englebert2021,Tusnin2020}. Both methods create an effective periodic potential, which provides an additional degree of freedom for controlling spatio-temporal dynamics and emerging states. Together with the stabilization of chaotic states \cite{Lobanov2016}, the potential may lead to the emergence of chimera-like states \cite{nielsen_nonlinear_2021,Tusnin2020}.
Furthermore, a modulated background provides different advantages, such as enhancing the pump-to-soliton conversion efficiency \cite{Erkintalo2022}, and providing additional deterministic routes for DKSs generation, without undergoing a spatio-temporal chaotic phase \cite{Taheri2015}.



In this letter, we theoretically show that a parabolic potential in time plays a key role on the  stability of DKSs and other spatio-temporal dissipative structures emerging in a dispersive Kerr resonator with anomalous dispersion. The parabolic potential approximates a periodic (e.g., sinusoidal) potential around the center of the DKS. Specifically, we find that, for low pump values, the potential stabilizes oscillatory and chaotic dynamics in favor of static DKS. As the pump power grows larger, the potential induces the appearance of asymmetric breathers and {\it chaoticons} (i.e. chimera-like states) \cite{verschueren_chaoticon_2014}, where the background field state coexists with {\color{black}a spatio-temporal localized chaotic state.} Moreover, chaoticons coexist with single-peak DKSs, and form a hysteresis loop. To support our findings, we carry out a systematic bifurcation analysis, which establishes the connection with the multimodal structure of the potential. 

In the mean-field approximation, the coherently driven and phase modulated passive cavity is described by the equation 
\begin{equation} 
\partial_t A = i \partial_\tau^2 A - iC\tau^2A 
+ i|A|^2A  - (1+i\delta)A+ P,
\label{eq1}
\end{equation}
where $A(\tau,t)$ is the slowly varying envelope of electric field, and $\tau$, $t$ are the fast and slow time, respectively\cite{haelterman_dissipative_1992}. The term $\partial^2_\tau$ is second-order anomalous chromatic dispersion, 
$\delta$ is the phase detuning, $P$ is the driving pump field amplitude, and
the linear loss coefficient, without loss of generality, is fixed to $1$.
We introduce the parabolic temporal potential $C\tau^2$, where $C$ controls its curvature. Note that such type of potential describes a trap in Bose-Einstein condensates, and a transverse index profile in graded-index multimode fibers \cite{krupa2019multimode}. With the usual change of the meaning of the coordinates (e.g., modify time $\tau$ with space $x$), Eq.(1) also describes the spatial dynamics of one-dimensional (e.g., consider slab waveguides) driven nonlinear passive cavities with a graded refractive index \cite{HAELTERMAN1992343}. 
To study the dynamics of  Eq.~(\ref{eq1}), we perform both direct numerical simulations (DNSs) with a pseudo-spectral method,
and 
{\color{black} numerical path-continuation of the stationary solutions $A_s$ (i.e., $\partial_t A_s=0$) by using 
AUTO-07p \cite{Doedel2009}. The latter allows us to compute both stable and unstable steady-state solutions, which are not accessible otherwise. }



\begin{figure}[!t]
\centering
\includegraphics[scale=1]{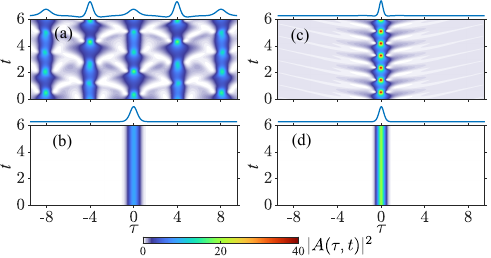}
\caption{Comparison of solutions without the potential $C=0$ in (a,c) and with the potential $C=1$ in (b,d), when the parameters are (a,b) $P=2.5$, $\delta=3$ and (c,d) $P=4.5$, $\delta=8$.  }
\label{fig1_stabilization}
\end{figure}

Figure~\ref{fig1_stabilization}(a,b) shows the dynamics of solutions of Eq.(1) in the absence of the potential $(C=0)$. The temporal evolution ($|A(\tau)|^2$ vs. $t$) of a chaotic Turing pattern, and its final state are shown in Fig.~\ref{fig1_stabilization}(a) for $(P,\delta) = (2.5,3)$.  Whereas {\color{black}Fig.~\ref{fig1_stabilization}(c)} shows a breather DKS for $(P,\delta)=(4.5,8)$. For this set of parameters, static DKSs are always unstable \cite{Leo2013}. When the potential is introduced $(C=1)$ [see Figs.~\ref{fig1_stabilization}(b) and \ref{fig1_stabilization}(d)], these dynamics are stabilized, leading to stationary DKSs.

\begin{figure}[!t]
\centering
\includegraphics[scale=0.94]{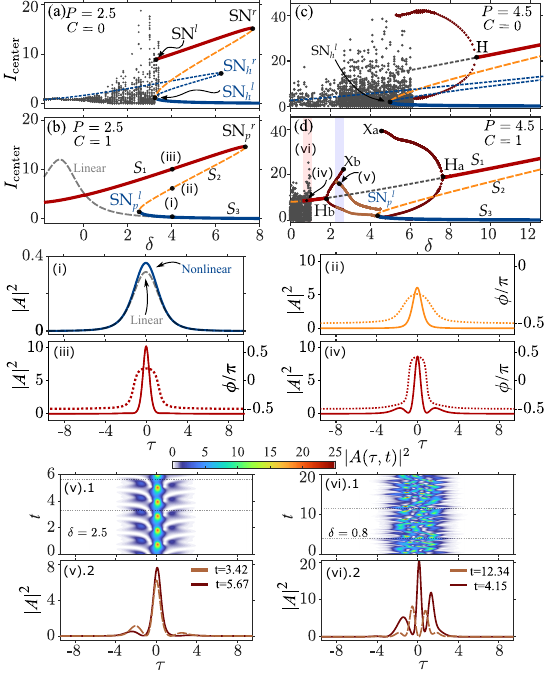}
\caption{Bifurcation diagrams showing $I_{\rm center}$ vs. $\delta$ for $(P,C)=(2.5,0)$ in (a), $(P,C)=(2.5,1)$ in (b), $(P,C)=(4.5,0)$ in (c), and $(P,C)=(4.5,1)$ in (d). 
{\color{black}The dashed gray line in (b) is the linear steady-state solutions, to be compared with with the nonlinear solution.}
Panels (i)-(iii) show the intensity $|A|^2$ (solid line) and phase $\phi$ (dashed line) of the three localized solutions $S_1$, $S_2$ and $S_3$, corresponding to the curves in panels (b,d); panel (iv) corresponds to the DKS solution in panel (d).
Panel (v).1 shows the evolution of a $\tau$-asymmetric state for $\delta=2.5$; panel (vi).1 shows a chaoticon for $\delta=0.8$, corresponding to panel (d)
[see \href{https://youtu.be/7fObrzTnA7c}{Visualization I}
 and \href{https://youtu.be/Vhz-yL41sB0}{Visualization II}
].
We plot the corresponding field powers at different slow times in panels (v).2 and (vi).2, respectively. 
}	
\label{fig2_bifur}
\end{figure}

In order to understand the mechanism for this stabilization, we perform a bifurcation analysis of
the DKSs, with and without a temporal potential. 
These results are illustrated in Fig.~\ref{fig2_bifur}, using $I_{\rm center}\equiv|A(0)|^2$ vs. $\delta$. {\color{black}
We set  $P=2.5$ in Figs.~\ref{fig2_bifur}(a,b), $P=4.5$ in Figs.~\ref{fig2_bifur}(c,d), $C=0$ in Figs.~\ref{fig2_bifur}(a,c), and $C=1$ in  Figs.~\ref{fig2_bifur}(b,d). Stable and unstable steady-state branches are computed by path-continuation algorithms; dynamical states are calculated by DNSs.  }


Figure~\ref{fig2_bifur}(a) shows the bifurcation diagram in the absence of potential, with $P=2.5$. The blue curve corresponds to the continuous-wave (CW) state of Eq.(1). The CW state is stable until the saddle-node (SN) bifurcation SN$_{h}^l$, where it becomes unstable [see dashed blue lines]. The DKS bifurcates from SN$_{h}^l$ with a small amplitude, and it remains unstable [see orange dashed lines] \cite{parra-rivas_bifurcation_2018-1}. By increasing $\delta$, the DKS eventually stabilizes at SN$^r$, and it retains stability until reaching SN$^l$ [see solid red line]. These solitons have a non-zero background (corresponding to the CW state), and their localized profile can be approximated by a sech-shape \cite{wabnitz:93}. When decreasing $\delta$ below SN$_h^l$, the DKS background becomes unstable, leading to chaotic Turing pattern states, such as in the example shown in Fig.~\ref{fig1_stabilization} for $(\delta,P)=(3,2.5)$. The peak intensity values of these states are plotted by gray dots in Fig.~\ref{fig2_bifur}(a).


This scenario drastically changes in the presence of the parabolic potential [ see Fig.~\ref{fig2_bifur}(b)]. Now, the CW state diagram merges with the solution branches corresponding to the DKS, leading to the single curve of Fig.~\ref{fig2_bifur}(b). Each branch on this curve corresponds to {\color{black}a localized state}, as depicted in Fig.~\ref{fig2_bifur}(i)-(iii), and branches are interconnected through the SN bifurcations SN$_{p}^{l,r}$. 


The $S_3$ state plotted in Fig.~\ref{fig2_bifur}(i) is a small-amplitude localized pulse, which corresponds to the deformation of the CW state, owing to the presence of the potential. This state extends until SN$_{p}^l$, where it connects to the unstable state $S_2$ [see Fig.~\ref{fig2_bifur}(ii)]. At SN$_{p}^r$, the latter leads to the DKS state $S_1$ [see Fig.~\ref{fig2_bifur}(iii)]. {\color{black} $S_1$ rests on the basal state $S_3$, and is asymptotically connected to a zero intensity background.} For $C=1$, $S_1$ extends to {\color{black} negative $\delta$} and it is stable. Between SN$_{p}^l$ and SN$_{p}^r$, {\color{black} the localized states $S_1$ and $S_3$ coexist for the same range of parameters, and are both stable. They can be easily excited by a Gaussian function of the form $A(\tau) = h\exp(-(\tau/r)^2/2)$, with $h$ and $r$ taking different values. In the absence of nonlinearity, bistability disappears: the linear resonance of Eq.(1) is shown by the dashed gray line in Fig.~\ref{fig2_bifur}(b). The linear solution, obtained by removing Kerr term, is also plotted in Fig.~2(i).
Therefore, the solution branch $S_3$ represents a nonlinear deformation of the linear steady-state solution. 
}

As we have anticipated with Fig.~\ref{fig1_stabilization}, the stabilization of the dynamics of solutions to Eq.(1) occurs for different values of $P$. To support this, we computed the bifurcation diagram associated with a single-peak soliton for $P=4.5$, with and without the potential. This situation is depicted in Fig.~\ref{fig2_bifur}(c,d). Figure~\ref{fig2_bifur}(c) shows the bifurcation diagram for $C=0$. By increasing $P$, the role of nonlinearity grows larger, which further tilts the resonance [see the blue lines]. The DKS solution branches preserve the morphology depicted in Fig.~\ref{fig2_bifur}(a), although now their range of existence has increased. In this regime, the top DKSs branch undergoes a Hopf bifurcation (H), where the soliton becomes unstable in favor of breathing states. The minimal and maximal value of $I_{\rm center}$ of these breathers are plotted by means of brown dots. These breather states are similar to that depicted in Fig.~\ref{fig1_stabilization}(c), and their oscillation amplitude grows larger with decreasing values of $\delta$. Eventually, when SN$_h^l$ is crossed, the stable CW state disappears, and spatio-temporal chaos (STC) develops. 
Note that STC extends the breather over SN$_h^l$, and coexists with DKSs or breathers [see Fig.~\ref{fig2_bifur}(a,c)].

For $C=1$ [see Fig.~\ref{fig2_bifur}(d)], STC is suppressed by the potential, in favor of either static DKS or regular oscillatory states. The DKS $S_1$ enlarges its stability region, which now extends to H$_a$, where $\delta_{{\rm H}_a}\ll\delta_{{\rm H}}$. Once H$_a$ is crossed, a $\tau$-symmetric breather arises supercritically, and it increases its oscillation amplitude with decreasing values of $\delta$. Eventually, this stable breather disappears, possibly in a fold of cycles at $X_a$. By decreasing $\delta$ below this point, the system develops $\tau$-asymmetric breathers, such as the one which is depicted in Fig.~\ref{fig2_bifur}(v) for $(\delta,P)=(2.5,4.5)$.
A special feature of these states is the different evolution of their leading and trailing tails [See \href{https://youtu.be/7fObrzTnA7c}{Visualization I}]. The extrema of these states, at $\tau=0$, are depicted by using brown dots in Fig.~\ref{fig2_bifur}(d). Decreasing $\delta$ further, the asymmetric breather branch meets with a symmetric one [see dark red branch] and disappears. On the right, the latter persists until reaching $X_b$. On the left, the $\tau$-symmetric breather decreases its amplitude, until it dies out at the Hopf bifurcation H$_b$. Note the presence of a bistability region between the symmetric and asymmetric breathers [see the light blue shadowed area in Fig.~\ref{fig2_bifur}(d)]. For $\delta<\delta_{{\rm H}_b}$, DKSs exist [such as the one shown in Fig.~\ref{fig2_bifur}(iv)], although they lose stability once more around $\delta\approx0.5$. 
After this, the solution of Eq.(1) evolves into a very complex spatio-temporal state, such as the one shown in Fig.~\ref{fig2_bifur}(vi). It consists of a portion of STC which is localized around the center of the temporal domain, owing to the presence of the potential which acts as a trap, thus confining the STCs. 
{\color{black} Such type of chaotic pulse was named chaoticon by Vershueren {\it et al.} \cite{verschueren_chaoticon_2014}, although it is also known as chimera state in other works \cite{nielsen_nonlinear_2021,Tusnin2020}. Chaoticons and DKSs [see respectively Fig.~\ref{fig2_bifur}(vi, iv) and \href{https://youtu.be/Vhz-yL41sB0}{Visualization II}
] coexist within a given $\delta$-range [see the pink shadowed area in Fig.~\ref{fig2_bifur}(d)].}

\begin{figure}[t]
\centering
\includegraphics[scale=1]{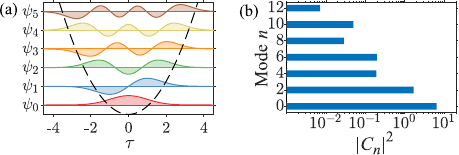}
\caption{(a) shows the first six HG linear modes $\psi_0,\cdots, \psi_5$ of Eq.(1) and the parabolic potential [see dashed line], for comparison. (b) illustrates the mode energy distribution associated with the $S_1$ DKS shown in Fig.~\ref{fig2_bifur}(iii) for $(\delta,P)=(4,2.5)$, where $|C_n|^2$ represents the energy of the mode $n$.}
\label{fig3_modes0}
\end{figure}

The previously described states can be analyzed in terms of a mode decomposition method.
The parabolic potential introduces boundary conditions for the fields, which translate in a finite number of eigenmodes. Therefore, solutions of Eq.(1) in Fig.~\ref{fig2_bifur} can be decomposed and analyzed in terms of linear
eigenmodes. {\color{black} This provides essential information regarding the underlying nonlinear mode interactions and the global dynamics of Eq.(1).} 
The linear eigenmodes obey the equation  
\begin{equation}
\partial_t A =i \partial_\tau^2 A - i C\tau^2A,
\label{eq_quantum}
\end{equation}
which also describes a quantum mechanical harmonic oscillator. Note that a similar equation was also used for describing the dynamics of mode-locked nanolasers \cite{Sun2019,Sun2020}.
As it is well-known, the eigenmodes of Eq.~(\ref{eq_quantum}) are the Hermite-Gaussian (HG) family. The lowest order six modes are plotted in Fig.~\ref{fig3_modes0}(a). 
The neglected terms in Eq.~(\ref{eq_quantum}), compared with Eq.~(\ref{eq1}), can be considered as small perturbations.
The field envelope $A(\tau,t)$ can be written as a linear superposition of HG modes $\psi_n(\tau)$ with equally spaced frequencies $2\sqrt{C}(n+1/2)$:
$A(\tau,t) = \sum_{n=0}^NC_n(t){ \psi}_n(\tau)$,
with $N$ being the total number of modes considered in the analysis. 
The mode coefficients are computed by projecting any solution on the linear modes, and read as
\begin{figure}[t]
\centering
\includegraphics[scale=1]{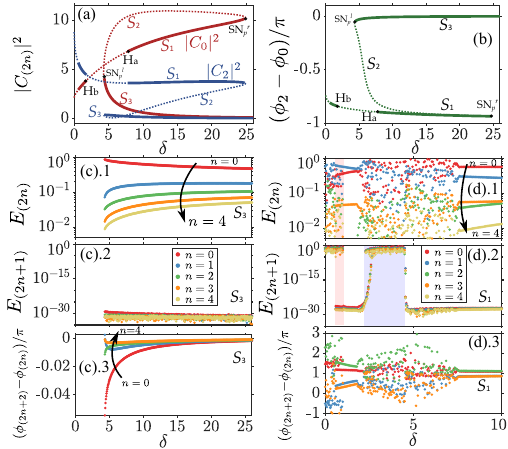}
\caption{Mode decomposition of the solutions in Fig.~\ref{fig2_bifur}(d). Mode energies $|C_0|^2$, $|C_2|^2$ in (a) and their phase difference in (b) as a function of $\delta$, where solid (dashed) lines represent stable (unstable) solutions.
Three panels in (c) plot  {\color{black} the normalized energies of even modes  $E_{(2n)}=|C_{(2n)}|^2/\sum|C_{(n)}|^2$, odd modes $E_{(2n+1)}$}, and their phase difference $(\phi_{(2n+2)}-\phi_{(2n)})/\pi$ vs. $\delta$, for $S_3$ solutions in Fig.~\ref{fig2_bifur}(d). The same quantities of the solutions $S_1$ of Fig.~\ref{fig2_bifur}(d) are shown in three panels in (d). {\color{black} Chaoticon/DKSs coexistence (asymmetric breathers) is represented by a pink (blue) shadowed area. }
    }
\label{fig4_modes}
\end{figure}
$C_n(t) = \int_{-\infty}^{\infty}A(\tau,t)\psi_n(\tau)\dd \tau=|C_n(t)|\exp{\mathrm{i}\phi_n(t)}$,
where $|C_n(t)|^2$ represents the energy of mode $n$ at time $t$, and $\phi_n(t)$ is its phase. 
In Fig.~\ref{fig3_modes0}(b), we plot the mode energy $|C_n|^2$ distribution associated with the DKS state $S_1$ in Fig.~\ref{fig2_bifur}(iii). Note that the energies of asymmetric modes ($n=1,3,5,\cdots$) are zero, because the state has a symmetric temporal distribution. $S_2$ and $S_3$ have a similar mode distribution [not shown here]. 

The bifurcation structure shown in Fig.~\ref{fig2_bifur}(d) ($P=4.5$) can be revisited by projecting the different DKS branches on HG modes. This projection is depicted in Fig.~\ref{fig4_modes}(a), where we plot the mode energy $|C_0|^2$ and $|C_2|^2$ vs. $\delta$. 
Following this diagram, we can see how the $\psi_{0,2}$-mode composition varies along the solution branches $S_1$, $S_2$ and $S_3$.

{\color{black}
Interestingly, we found that $S_1$ and $S_3$ represent mode-locked states.
This is shown in Fig.~\ref{fig4_modes}(b), where we show the phase difference between modes $\psi_0$ and $\psi_2$ along the bifurcation diagrams of Fig.~\ref{fig4_modes}(a). 
The phase difference between adjacent symmetric modes is $-\pi$ in $S_1$ [other higher-order modes are not shown here]: these modes are locked in anti-phase, in contrast with the in-phase mode locking that occurs for $S_3$ states.
Whereas unstable states $S_2$ undergo large phase changes between the two modes. 
Useful insight in the localized states $S_1$ and $S_3$ is gained by analyzing the variation of their mode components with cavity detuning $\delta$.
For the $S_3$ branch, the energy fraction of higher-order modes $E_{(2n)}=|C_{(2n)}|^2/\sum|C_{(n)}|^2$ grows larger as $\delta$ increases [see Fig.~\ref{fig4_modes}(c).1]. This leads to temporal broadening of the $S_3$ states, because of the longer temporal duration of higher-order modes. 
On the other hand, for $S_1$ the fundamental mode energy $|C_0|^2$ grows significantly larger with $\delta$ [see Fig.~\ref{fig4_modes}(a)]. This results in a cavity soliton with progressively higher intensity and narrower temporal duration. 
For both states, the phase difference between adjacent even modes approaches zero when $\delta$ increases [see Fig.~\ref{fig4_modes}(c).3 and Fig.~\ref{fig4_modes}(d).3]. This shows that stronger locking occurs for these modes with increasing $\delta$.
Furthermore, the mode decomposition analysis allows for making a clear distinction among the symmetric and asymmetric breathers, and chaoticons.
In the symmetric breather regime ($1.8<\delta<7.5$), 
only even mode energies [see Fig.~\ref{fig4_modes}(d).1] and phase differences [see Fig.~\ref{fig4_modes}(d).3] fluctuate. This is in contrast with the case of asymmetric breathers, which exhibits significant odd mode components [see Fig.~\ref{fig4_modes}(d).2]. These odd modes  contribute to the asymmetric breather evolution. When compared with asymmetric breathers, chaoticons exhibit much wider phase fluctuations [see Fig.~\ref{fig4_modes}(d).3].
}



In summary, by applying a bifurcation analysis, we revealed the emergence and stability of dissipative states for a coherently driven, passive nonlinear and dispersive cavity with a parabolic potential. The potential may stabilize complex spatio-temporal dynamics in favor of static DKSs, and leads to the coexistence of high and {\color{black}low amplitude localized states.} A particular feature of this system is that asymmetric breathers and chimera-like states (i.e., a chaoticons) may arise. The latter consist of localized spatio-temporal chaos, and appear due to potential trapping. By a modal decomposition analysis, we have shown that these states emerge from nonlinear interactions of asymmetric modes. The simple parabolic potential captures the essential dynamics introduced by synchronous phase modulation 
and permits to gain useful physical insight.

\section*{Acknowledgments}
This work was supported by European Research Council (740355), Marie Sklodowska-Curie Actions (101064614,101023717), Ministero dell’Istruzione, dell’Università e della Ricerca (R18SPB8227).

\section*{Data availability}
 Data underlying the results presented in this paper are stored in the open sharing platform Zenodo and can be accessed directly through the link: \href{https://zenodo.org/record/7484479}{https://zenodo.org/record/7484479}
.

\bibliography{main}

\end{document}